\documentstyle[twocolumn,prb,aps]{revtex}
\def\be{\begin{equation}}
\def\ee{\end{equation}}
\def\bc{\begin{center}}
\def\ec{\end{center}}
\begin{document}
\title{Stability and diffusion of surface clusters}
\author{T. M\"uller and W. Selke}
\address{
Institut f\"ur Theoretische Physik, Technische Hochschule,
D--52056 Aachen, Germany}

\maketitle

\begin{abstract}
Using kinetic Monte Carlo simulations and a bond--counting
ansatz, thermal stability and diffusion of an adatom island
on a crystal surface are studied. At low temperatures, the
diffusion constant $D$ is
found to decrease for a wide range of island sizes like 
$D \propto N^{-\alpha}$, where $\alpha$ is close to one, $N$
being the number of adatoms in the cluster. By
heating up the surface, the system undergoes a phase transition
above which the island disappears. Characteristics of that
transition are discussed.
\end{abstract}

\vspace{0.2cm}
\noindent

 
\vspace{1cm}

\section{Introduction}

Stability and dynamics of adatom as well as vacancy islands of
monoatomic height on crystal surfaces have been studied 
intensively during the last years, both experimentally
and theoretically [1-12]. In particular, the equilibrium
island size at low temperatures [10] as well as the
decay of clusters
in the presence of surface steps or larger
islands ('Ostwald ripening') [9,11-12] have been analysed. Different
microscopic mechanisms have been discussed to explain the
size dependence of the diffusion constant $D$ characterising
the motion of the center of mass of an equilibrated
island. Typically, $D$ decreases with the number of adatoms $N$
in the cluster in the form of a
power--law, $D \propto N^{-\alpha}$, with $\alpha$ depending on the 
mechanism driving the island motion [1-8].

In this article, we consider adatom islands 
on a (100) surface of a simple cubic crystal with energy barriers
for jumps of the adatoms to neighbouring surface sites given
by isotropic nearest--neighbour bond energies. Equilibrium and
dynamic properties are computed by using kinetic Monte Carlo
techniques [13]. Specificly, for a single
island in equilibrium with the 'gas' of adatoms
on the surrounding terrace a phase transition is observed at
a critical temperature $T_c$, above which the island tends to
disappear. This aspect seems to have been overlooked in
previous investigations. A study on the
equilibrium cluster size in a related Ising model
had been performed already several years ago [14], but it had 
been limited to low temperatures. Quite recently, the
thermal disintegration of a cluster was
discussed [15], however, for systems without conservation of
the number of particles. In addition, we
monitor the motion of the 
equilibrated adatom island well below $T_c$, to estimate the
value of $\alpha$ in
the nearest neighbour isotropic bond--counting case.

The article is organized accordingly. In the next Section, we
introduce model and method. We then present results on the
characteristics of the
phase transition at which the island disappears, followed by
a discussion of the diffusive motion of the cluster at
low temperatures. We conclude with a short summary.
 
\section{Model and method}

Adatoms on a square lattice may be constrained to a single layer, with
the occupation variable $n_i$ at surface site $i$ being either
1 or 0. To jump to an empty neighbouring site, the adatom has
to overcome an energy barrier $E_a$. Using a bond--counting ansatz, that
energy may be written in the form
 
\be
E_a= E_0 + n E_b
\ee

\noindent
where $E_0$ is the activation energy for free diffusion of the adatom
on a locally perfect surface, $n$ is the number of occupied
neighbouring sites, $n$= 0,1,2, or 3, and $E_b$ is the bond
energy. Of course, the ansatz is not expected to give a realistic
description of a specific material. However, it is useful in
identifying generic properties of islands, as will be discussed
below.

The energy barrier (1) corresponds to the Hamiltonian (apart from
a constant)  

\be
{\cal H} = -E_b~ {\sum\limits_{i, j}} n_i n_j 
\ee

\noindent
where the sum runs over neighbouring sites $i$ and $j$. It is 
interesting to note that the Hamiltonian may be transcribed to a
nearest--neighbour Ising model [11],

\be
{\cal H_I} = -E_b/4~ {\sum\limits_{i, j}} S_i S_j  +constant
\ee

\noindent
with the Ising spin $S_i= \pm 1$ being related to the occupation
variable $n_i$ by $S_i= 2 n_i -1$. The 
conservation of the number of adatoms on the surface corresponds to the
conservation of the magnetization in the Ising model.

We considered square surfaces with $L\times L$ sites and 
$M\times M$ adatoms, i.e. with a coverage $\theta= M^2/L^2$. Usually, full
periodic boundary conditions were used, to reduce boundary effects. However, to
compare with possible experiments, we also applied free boundary
conditions as encountered for a terrace of $L^2$ sites with
large reflecting energy barriers at the descending steps bordering
the terrace (large Schwoebel--Ehrlich barriers).

In the simulations, $L$ ranged from 50 to 1000, and $M$ from 5 to 50, with
the coverage $\theta$ varying between $4\times 10^{-4}$ and
$4\times 10^{-2}$. Obviously, at those coverages the adatoms form a compact
island at low temperatures due to the attractive interactions, see (1). A
typical configuration is shown in Fig. 1.

The kinetic Monte Carlo simulations were performed in the standard
way [13,16], based on jump probabilities for the adatoms to
neighbouring sites $\propto \exp(- E_a/k_BT)$, with $k_B$ being
the Boltzmann constant and $T$ the temperature. The time may be
measured in units of trial jumps per adatom (MCA). Other commonly
used time scales, invoking, e.g., a microscopic attempt frequency $\nu$, are
linearly related to that unit. At low temperatures, the efficient
algorithm of Bortz, Kalos, and Lebowitz [17], BKL, was implemented.

To study stability and dynamics of the island, several quantities were
computed. Among others, we recorded the distribution of clusters (as
usual, a $s$-cluster is formed by $s$ adatoms connected by occupied
neighbouring sites), especially the fraction of adatoms in the
largest cluster, i.e. the 'reduced island size',

\be
n_{max}= N_{max}/M^2 
\ee

\noindent
where $N_{max}$ is the number of adatoms in the largest
cluster. Furthermore, the fluctuations of that island size, the density of
adatoms, the energy $E$, see (2), and the fluctuations of the energy
were monitored. To analyse the motion of the island, we determined
the time evolution of the position of its center of
mass, see Section IV.

\section{Thermal stability}

In the ground state, $T= 0$, the $M^2$ adatoms form a compact square
cluster, at sufficiently low coverage $\theta$ (otherwise, a stripe
of adatoms will minimize the energy). At non--zero temperatures, adatoms
may detach from the island, leading to a dynamic equilibrium of
the rounded cluster and the 'gas' of adatoms on the
terrace, see Fig. 1. Of course, the size of the island is expected 
to shrink as the temperature increases, as demonstrated in
Figs. 2 and 3.

For reasons of simplicity, we assume equal energy barriers in the
bond--counting ansatz, (1), i.e. $E_0= E_b$. Starting the simulations
with a square cluster of $M^2$ adatoms, the time 
evolution of the island size, $n_{max}(t)$, at various temperatures
$k_BT/E_b$, is shown in Fig. 2. Obviously, equilibration may be
rather slow, as observed before in the related Ising model [14]. Discarding
the initial relaxation, the thermal equilibrium value for the reduced
island size, $n_m(T)= \left< n_{max} \right>$, is obtained
by averaging over the
subsequent, possibly strongly fluctuating, see Fig. 2, simulational data.
The resulting temperature dependent island size is depicted
in Fig. 3, at coverage $\theta= 0.04$ with $M$= 25 and 50 (hence, $L$= 125
and 250).

The drastic decrease of $n_m(T)$, both for periodic and free
boundary conditions, in a narrow range of temperatures suggests a phase
transition at the critical temperature $T_c$ in the thermodynamic
limit, $L, M \longrightarrow \infty$, at constant
coverage $\theta= M^2/L^2$. The reduced island size $n_m(T)$ may then
be interpreted as the order parameter, vanishing at $T \ge T_c$. In
the high--temperature phase, the island disappears in the gas of 
adatoms. Indeed, finite--size analyses of the Monte Carlo
data, at fixed coverages, allow to locate the
phase transition: For example, at $\theta= 0.01$, the
turning point, $T_m(M)$, of the temperature dependent island
size $n_m(T)$, see Fig. 3, is
found to shift for large clusters $M$ approximately
like $T_c - T_m(M) \propto 1/M$, and at
temperatures above $T_c$, $n_m(T)$ goes to zero inversely proportional
to $M^2$ (for details, see Ref. 18). Moreover, the transition
is signalled by a pronounced maximum in the temperature derivative
of the energy close to $T_m$, similarly to that of the fluctuations
in the island size and the energy. All these quantities seem to
show singular behaviour in the thermodynamic limit.

In general, a phase diagram in the temperature--coverage 
($k_BT/E_b, \theta$) plane may be determined, with a single
large cluster or, at $\theta \ge 1/4$, a stripe of adatoms characterising
the low--temperature phase. Because of the transcription to the
Ising model, (3), $T_c$ is known exactly at $\theta$= 1/2, namely
$k_BT/E_b= 1/(2\ln(\sqrt 2 +1))$. Certainly, the transition temperature
tends to decrease with decreasing coverage. For instance, we
estimated, at $\theta$= 0.04, 0.01, and 0.0016 the critical
temperatures $k_BT/E_b \approx$ 0.49,
0.38, and 0.28, respectively.

Approaching the phase transition from below, the largest cluster
becomes more and more ramified and may dissociate quite easily, while other
groups of adatoms may coalesce forming a new largest cluster, as
one may readily observe in the simulations. Accordingly, the
island size $n_m(T)$ as well as the energy may fluctuate strongly, and
good statistics is needed to get reliable equilibrium values in
the critical region, and to quantify the asymptotics. In particular, we
analysed the order parameter $n_m(T)$, as $T \longrightarrow T_c$. Fitting
the simulational data to a power--law 

\be
n_m(T) \propto (T_c - T) ^{\beta} 
\ee

\noindent
we obtained, from extensive simulations at
coverages $0.0016 \le \theta \le 0.04$, $\beta = 0.45 \pm 0.01$, being (if
at all) only fairly weakly
dependent on the coverage. The rather large error bar
reflects, especially, uncertainties in extrapolating the
estimates to the thermodynamic limit. Actually, we computed
an effective exponent
$\beta_{eff}$= d$\ln (n_m)/$d$\ln (t)$, where  $t= \mid T_c - T\mid$ or
$\mid T_m - T\mid$, leading to upper and lower bounds for
the critical exponent $\beta$, which is
approached in the limit $L, M \longrightarrow \infty$ and
$t \longrightarrow 0$.

The standard description of a cluster in equilibrium with a gas
is based on the Gibbs--Thomson formula [19]. The equilibrium
density $\rho$ of the gas of adatoms in coexistence with a circular
island of radius $R$ is then [11,12]

\be
\rho (R)= \rho_s \exp[\gamma /(R\rho_i k_BT)] 
\ee

\noindent
where $\gamma$ is the free energy per unit length of the island
edge, $\rho_i$ is the density of the island, and $\rho_s(T)$ denotes
the density of a gas of adatoms in the presence of a straight
step. Modified formulae have been discussed [11,14] to
include, e.g., interactions between the adatoms on the
terrace. However, no attempt
is known to us to extend the description into the critical
region, where the island tends to disappear, $\rho \approx \rho_i$.

In our simulations, we confirmed that the density of adatoms $\rho$
is, indeed, constant, when exceeding a critical distance from
the center of the island. Moreover, at low temperatures and
coverages, the Gibbs--Thomson formula (6) predicts an
approximately logarithmic dependence of the characteristic
temperature $T_x$, with $n_m(T_x)= x$, on the coverage $\theta$

\be
k_BT_x \propto -1/\ln [\theta (1-x)] 
\ee

\noindent
fixing the number of adatoms $M^2$ and varying the surface
size $L^2$. To derive (7), one may use the low--temperature relation [12]
$\rho_s \propto \exp (-E_{ad}/k_BT)$, where $E_{ad}$ is the energy to
detach an atom from the step. Indeed, for example, our simulational data
at $M= 10$, $100 \le L \le 1000$, and $x= 1/2$ agree well with (7) [18].

\section{Diffusion}

The exchange of adatoms between the island and the surrounding terrace
gas leads to a diffusive motion of the equilibrated island. The corresponding
diffusion constant $D$ follows from the fluctuations in the position
of the center of mass of the cluster, ${\bf r}_{cm}(t)$,

\be
D= <({\bf r}_{cm}(t)-{\bf r}_{cm}(0))^2>/(4t) 
\ee

\noindent
where the brackets denote the equilibrium average.

From previous studies [1-8], one expects that $D$ depends on the
number of adatoms in the island, $N= \left< N_{max} \right>$, at
least asymptotically for large values of $N$, as

\be
D \propto N^{-\alpha} 
\ee

\noindent
with the value of $\alpha$ characterising the dominat mechanism
of exchange between the cluster and the gas, discriminating, e.g., 
island edge or periphery diffusion, terrace diffusion and 
evaporation--condensation kinetics (a similar classification holds for
step fluctuations [20,21]), see below.

To check the diffusive character of the island motion, (8), and to
determine the characteristic exponent $\alpha$, we performed 
kinetic Monte Carlo simulations, using the bond--counting ansatz, (1), with
$E_0$=0, thereby speeding up the dynamic processes. The time
may be measured in units of seconds. Invoking the microscopic attempt
frequency $\nu$, one trial jump per adatom, 1 MCA, corresponds then
to 1/(4$\nu$) seconds [13,16]. We chose $\nu= 10^{11}$ Hz. The lattice
constant of the square surface was set equal to one. To compute 
the fluctuations in the position of the center of mass of the
island, (8), we first equilibrated the system, before avering over
an ensemble of initial times as well as an ensemble of (up to 1000)
realizations. The simulations were done at low temperatures, well
below $T_c$, namely $k_BT/E_b= 0.2$ and 0.28, for islands of
sizes $N$ ranging from about 20 to about 700. The coverage was
fixed, $\theta= 0.01$. We applied here the BKL algorithm in our extensive
simulations.

The positional
fluctuations, $\left<({\bf r}_{cm}(t)- {\bf r}_{cm}(0))^2\right>$, are
found to increase, indeed, linearly in
time, even at early times. The
diffusion constant $D$ is then readily obtained from linear
regression. The resulting size dependent $D(N)$ at $k_BT/E_b$=0.28
is shown in Fig. 4. Discarding the smallest island size, the
characteristic exponent $\alpha$ is estimated
to be, on average, $\alpha= 1.02 \pm0.03$. At $k_BT/E_b$= 0.2, the
value of $\alpha$ seems to be consistent
with $\alpha=1$ as well ($\alpha= 1.04 \pm 0.04$), for
island sizes N ranging from 20 to 700.

From elementary geometry and energy considerations [14] and from
a Langevin theory [2], $\alpha$=1 may be argued to correspond
to an island motion driven by terrace diffusion, where the
adatom emitted by the island diffuses as a random walker on the 
terrace before attaching again (or a vacancy may diffuse
through the island).

In principle, other mechanisms may compete, in
particular, random detachments and attachments of adatoms
at the island edge (evaporation--condensation kinetics) or diffusion of
the atoms along the island edge (periphery diffusion), leading to
$\alpha$= 1/2 and 3/2, respectively [2,14]. In addition, periphery
diffusion may be hindered by corners and kinks, modifying, possibly, the
value of $\alpha$ [4,6]. Among
othres, details
of the shape of the islands as well as the activation energies
are expected to determine which mechanism dominates. In general, it
is reasonable to consider an effective characteristic exponent
$\alpha_{eff}$= -d$\ln D$/d$\ln N$, which may depend on island size
$N$ and temperature $T$, as observed in experiments and simulations.

In our case, Fig. 4 indicates an increase of $\alpha_{eff}$ from higher
values at small island sizes
towards $\alpha \approx 1$ at $N\approx 30-40$. Of course, another crossover
at cluster sizes exceeding those we studied, $N \approx 700$, cannot be
excluded. Actually, simulations on a related Ising model, applying
Kawasaki dynamics, were interpreted as providing
evidence for a crossover from periphery diffusion to 
evaporation--condensation kinetics, studying somewhat
smaller cluster sizes, $N \le 500$, and higher coverages, with
an average exponent $\alpha$ not far from one [14]. In
contrast to the previous study, we
determined $D(N)$ at fixed coverage. Obviously, our data do not show clearly
such a crossover, but it cannot be ruled out.

It seems interesting to note that $\alpha \approx 1$ has been found
in experiments on Ag(111), using scanning tunneling microscopy [1],
but other values of $\alpha$ have been reported for different
surfaces, reflecting
the above mentioned competing mechanisms.

\section{Summary}

Thermal stability and diffusion of an adatom island 
of monoatomic height on a square surface
have been studied, using a standard bond--counting ansatz for the
energy barriers and performing kinetic Monte Carlo simulations.

A phase transition has been identified, with the fraction of adatoms
in the largest cluster $n_m$ being the order parameter. $n_m(T)$ vanishes
on approach to the transition temperature $T_c$ as 
$n_m \propto \mid T_c -T \mid ^\beta$ with $\beta =0.45 \pm 0.10$ at
various small coverages $\theta$. In addition, the
energy as well as fluctuations
in the island size and in the energy exhibit singular behaviour
at $T_c$. Experimentally, such phenomena may be observed on a 
terrace bordered by descending steps with large Schwoebel--Ehrlich
barriers to allow for equilibration of the island with the
surrounding gas of adatoms on the terrace.

The diffusion constant $D$, describing the island
motion, decreases with the number of adatoms in the island $N$ like
$D \propto N^{-\alpha}$, with $\alpha$ being close to one for
an extended range of island sizes at temperatures well below
$T_c$ at $\theta= 0.01$. $\alpha= 1$ corresponds to the case where
the dominant mechanism driving the motion of the island is
terrace diffusion. An alternative interpretation invokes a crossover
from periphery diffusion to evaporation--condensation kinetics.

Of course, it would be interesting to investigate the robustness
of our findings against varying, especially, the activation
energies in a systematic way.

\vspace{1cm}

\noindent{\bf Acknowledgements}\\
 
It is a pleasure to thank M. Bisani, T. L. Einstein,
 G. Schulze Icking--Konert, and L. Verheij for useful
discussions and help.

 
\vspace{1.5cm}

\bc
{\Large \bf References}\\[2ex]
\ec
\begin{enumerate}
\item Morgenstern, K., Rosenfeld, G., Poelsma, B., Comsa, G.: Phys. Rev. Lett. {\bf 74}, 2058 (1995)
\item Khare, S. V., Bartelt, N. C., Einstein, T. L.: Phys. Rev. Lett. {\bf 75}, 2148(1995); Khare, S. V., Einstein, T. L.: Phys. Rev. B{\bf 54}, 11752 (1996)
\item Wang, S. C., Ehrlich, G.: Phys. Rev. Lett. {\bf 79}, 4234 (1997)
\item Pai, W. W.,  Swan, A. K., Zhang, Z., Wendelken, J. F.: Phys. Rev. Lett. {\bf 79}, 3210 (1997)
\item Soler, J. M.: Phys. Rev. B. {\bf 50}, 5578 (1994)
\item Bogicevic, A., Liu, S., Jacobsen, J., Lundqvist, B., Metiu, H. : Phys. Rev. B {\bf 57}, R9459 (1998)
\item Sholl, D. S., Skodje, R. T.: Phys. Rev. Lett. {\bf 75}, 3158 (1995)
\item Bitar, L., Serena, P.A., Garcia--Mochales, P., Garcia, N., Thien Binh, Vu : Surf. Sci. {\bf 339}, 221 (1995)
\item Levitan, B., Domany, E.: J. Stat. Phys. {\bf 93}, 501 (1998)
\item Krishnamachari, B., McLean, J., Cooper, B., Sethna, J.: Phys. Rev. B. {\bf 54}, 8899 (1996)
\item McLean, J. G., Krishnamachari, B., Peale, D. R., Chason, E., Sethna, J. P., Cooper, B. H.: Phys. Rev. B {\bf 55}, 1811 (1996)
\item Schulze Icking--Konert, G., Giesen, M., Ibach, H.: Surf. Sci. {\bf 398}, 37 (1998); Schulze Icking--Konert, G.: PhD thesis, RWTH Aachen (1998)
\item Kang, H. C., Weinberg, W. H.: J. Chem. Phys. {\bf 90}, 2824 (1989)
\item Binder, K., Kalos, M. H.: J. Stat. Phys. {\bf 22}, 363 (1980)
\item Lee, J., Novotny, M. A., Rikvold, P. A. : Phys. Rev. E {\bf 52}, 356 (1995)
\item Newman, M., Barkema, G: Monte Carlo Methods in Statistical Physics. Oxford: Oxford University Press 1998
\item Bortz, A. B., Kalos, M. H., Lebowitz, J. L.: J. Comp. Phys. {\bf 17}, 10 (1975)
\item M\"uller, T.: Diploma thesis, RWTH Aachen (1998)
\item Rowlinson, S., Widom, B.: Molecular Theory of Capillarity. Oxford: Clarendon Press 1982
\item Khare, S. V., Einstein, T. L.: Phys. Rev. B {\bf 57}, 4782 (1998)
\item Selke, W., Bisani, M.: In: Anomalous Diffusion: From Basis to Applications. (Lecture Notes in Physics, vol. 519) Pekalski, A. and Sznajd--Weron, K. (eds.), Berlin, Heidelberg, New York: Springer 1999, p.298 ff.
\end{enumerate}

\newpage
\bc
{\Large \bf Figure Captions}\\[2ex]
\ec
\begin{itemize}
\item[Fig. 1:] Island of monoatomic height in equilibrium
with the surrounding gas of adatoms.
 
\item[Fig. 2:] Time dependence of the island size $n_{max}(t)$ for
$M= 50$ and $L= 250$, at various temperatures $k_BT/E_b$. The
initial configuration of the simulations is a square cluster with
$M^2$ adatoms. The time is measured in trial jumps per adatom (MCA). 
 
\item[Fig. 3:] Temperature dependence of
the thermally averaged island size $n_m(T)$ for
surfaces with $125^2$ sites and $25^2$ adatoms (open circles: periodic
boundary conditions; full diamonds: reflecting boundaries) as well
as $250^2$ sites and $50^2$ adatoms (open squares: periodic 
boundary conditions). Error bars are smaller than symbol sizes.
 
\item[Fig. 4:] Logarithm of the diffusion constant $D$ vs.
logarithm of number of adatoms in
the largest cluster $N$ at coverage $\theta= 0.01$ and temperature
$k_BT/E_b$ =0.28. The dashed line corresponds to $\alpha= 1$. The
size of the surfaces ($L^2$) ranged from 
$50^2$ to $280^2$ in the simulations.

\end{itemize}
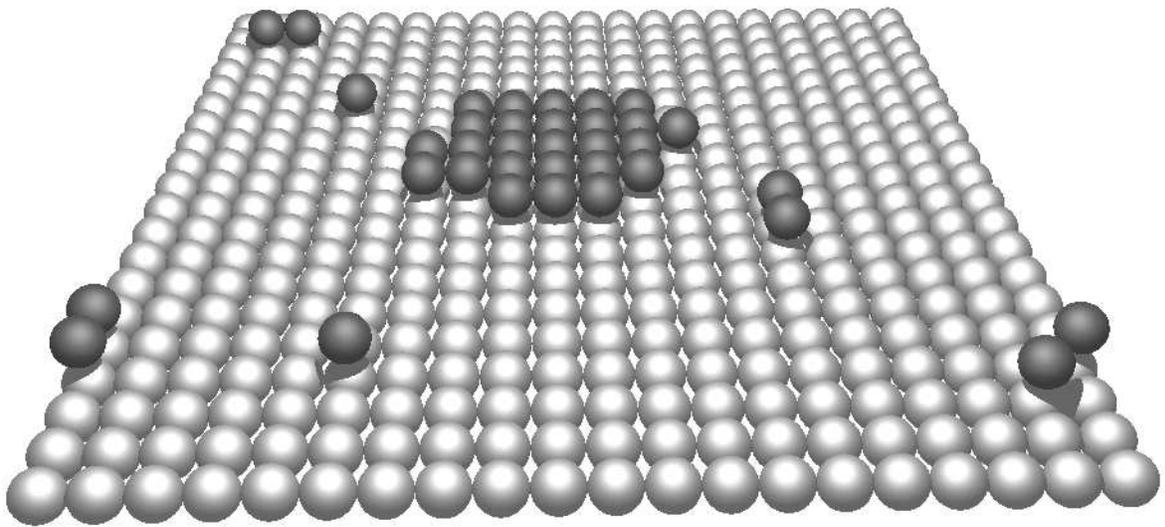
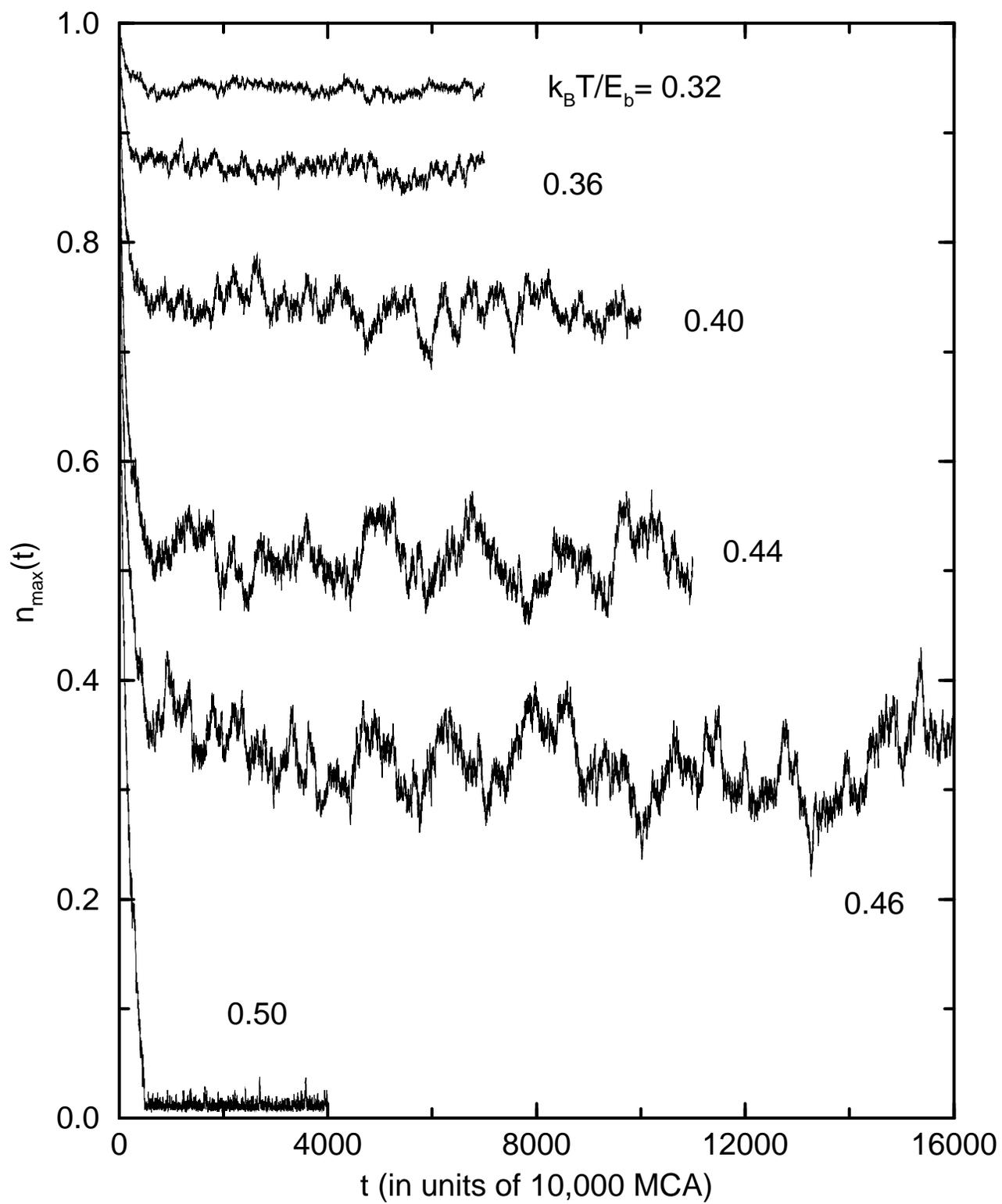
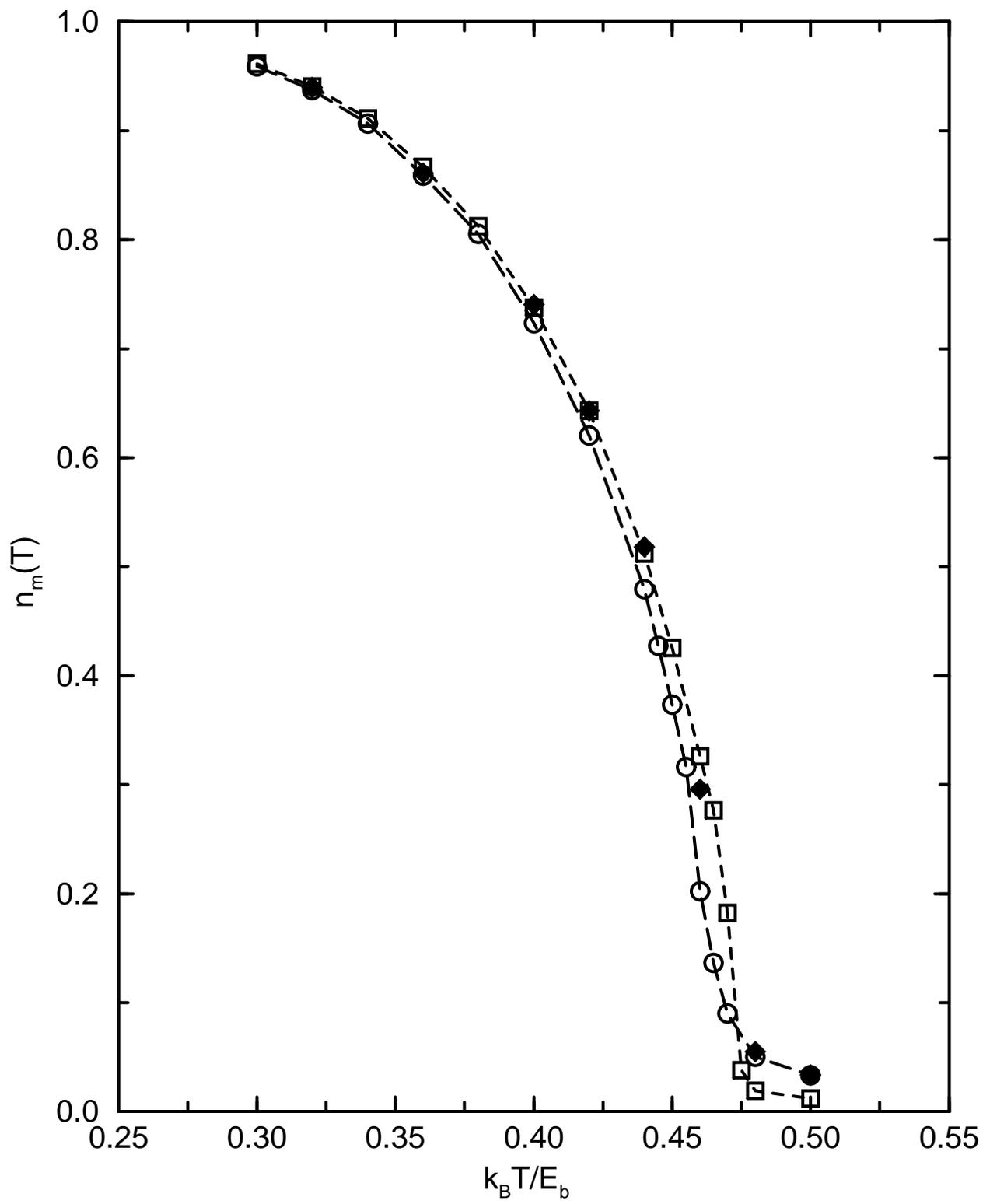
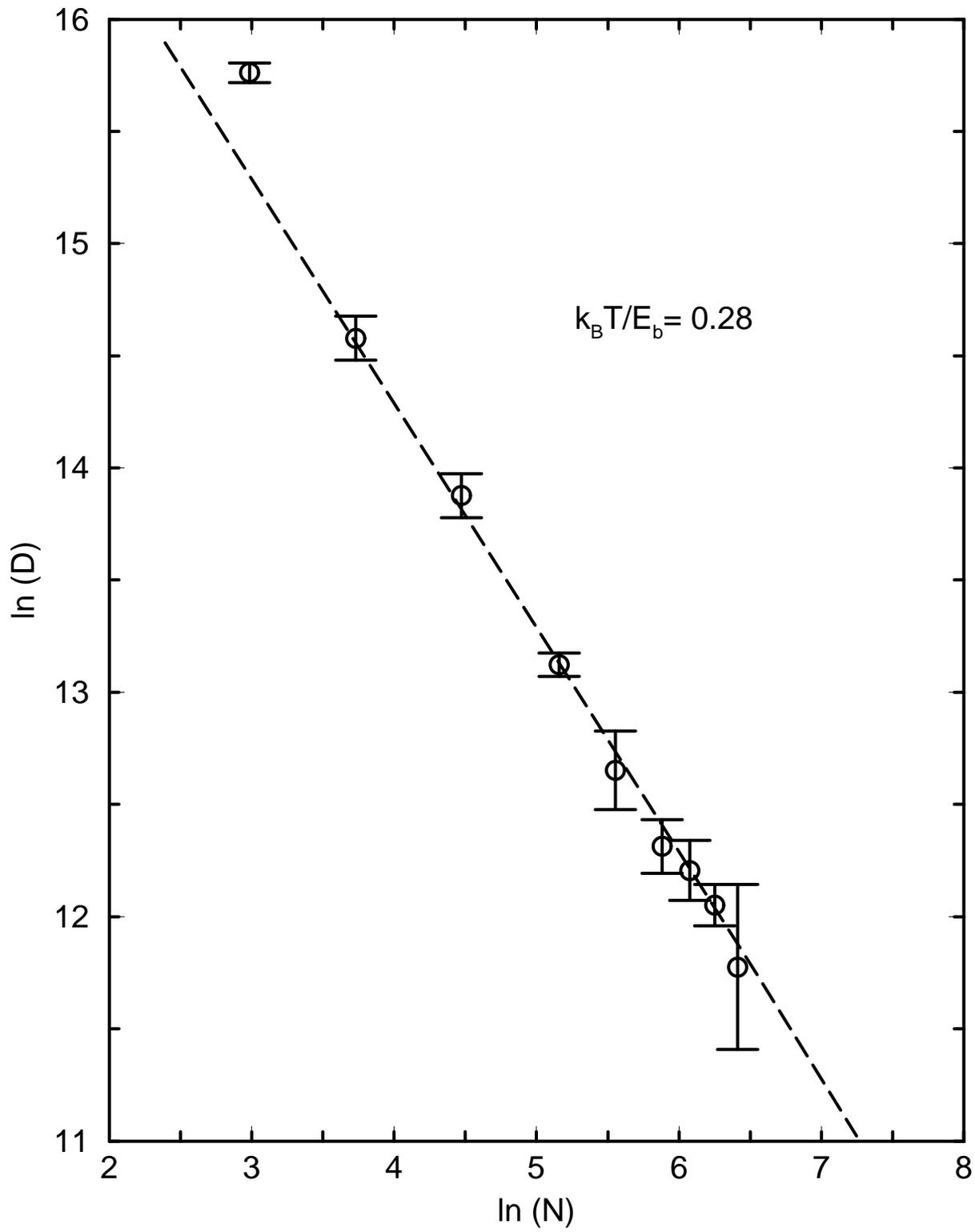
\end{document}